# *Federating and querying heterogeneous and distributed Web APIs and triple stores*

*Tarcisio Mendes de Farias*[1,2,*], *Christophe Dessimoz*[1,2], *Aaron Ayllon Benitez*[3], *Chen Yang*[4], *Jiao Long*[5] and *Ana-Claudia Sima*[1,2]

[1]SIB Swiss Institute of Bioinformatics, Lausanne, Switzerland
[2]University of Lausanne, Lausanne, Switzerland
[3]BASF Digital Solutions SL, Madrid, Spain
[4]BASF, Ghent, Belgium
[5]Ghent University, Ghent, Belgium

**ABSTRACT**

Today's international corporations such as BASF, a leading company in the crop protection industry, produce and consume more and more data that are often fragmented and accessible through Web APIs. In addition, part of the proprietary and public data of BASF's interest are stored in triple stores and accessible with the SPARQL query language. Homogenizing the data access modes and the underlying semantics of the data without modifying or replicating the original data sources become important requirements to achieve data integration and interoperability. In this work, we propose a federated data integration architecture within an industrial setup, that relies on an ontology-based data access method. Our performance evaluation in terms of query response time showed that most queries can be answered in under 1 second.

## 1 INTRODUCTION

Research and development (R&D) in agriculture at BASF, a leading company in crop protection industry, comprises seeds and traits research and breeding capacities; solutions that protect plants against fungal diseases, insect pests and weeds; improvements of soil management and plant health; and digital services for better crop management decisions. In this context, R&D produces many scientific datasets through various BASF groups worldwide. In addition, BASF also benefits from many sources of publicly available data, such as STRING, a protein-protein interaction database (Szklarczyk et al., 2021). Data integration and interoperability of public and internal datasets becomes crucial to achieving the holistic view required for innovative plant research.

Scientific data stored and produced by several stakeholder groups internally and externally to an organisation often require a deep understanding of the knowledge domain. The complexity of how the raw data are stored, processed, analysed, and structured by the scientific domain experts is often hidden through programmatic interfaces. As a result, data integration and interoperability would be hard to achieve at the physical data store layer by applying, for example, a data warehouse approach.

At BASF, the distinct niche scientific datasets are usually encapsulated with Web APIs over heterogeneous data stores (e.g. relational, hierarchical and property graph data models). Moreover, major private and public databases of BASF's interest natively model data with the Resource Description Framework (RDF) that are queried using SPARQL. Bridging the two data access modes, SPARQL and Web APIs, is of major importance for BASF. Indeed, by interoperating these data access modes, we reduce maintenance costs and stakeholders' coordination efforts, compared to a data warehouse solution (Sima et al., 2019). A non-exhaustive list of similar efforts are presented in (Michel et al., 2018) (Michel et al., 2019), that includes an application to biodiversity use cases.

To interoperate different access modes, we propose a federated data integration architecture that includes, besides native RDF stores, an ontology-based data access (OBDA) approach (Sima et al., 2019) over private and public Web APIs. We applied this architecture to successfully integrate and interoperate the BASF data. Due to confidentiality issues, in this paper, we focus on a use case only involving public data, namely, STRING and Orthologous MAtrix (OMA) databases. OMA contains data about evolutionary relationships among genes across species. Both are public databases that are, currently, in use and of interest to BASF. For example, BASF has its own in-house OMA database with its genomes of interest, that also includes proprietary data. As part of an OBDA solution, we have to develop several ontologies to semantically model the different Web API responses. Although most of them are private because they model proprietary data, in this work, we made public available the developed ontology related to STRING database Web APIs.

Our article is structured as follows. In Section 2, we describe the proposed architecture and its implementation, as well as a high-level introduction to the developed ontology for STRING. In Section 3, we discuss our public use case over OMA and STRING databases along with a short performance evaluation. Finally, we conclude with a discussion and outlook.

## 2 METHODS

To interoperate fragmented data sources accessible via Web APIs and SPARQL, we applied an OBDA approach over these APIs and we enabled a SPARQL 1.1 federation over the RDF native and "virtual" knowledge graphs (VKG). The knowledge graphs are called "virtual" when the data are not replicated but stay within the data sources and accessed at query time (Calvanese et al., 2021). In doing so, we promote data integration through mostly conjunctive queries at query execution time. These queries are federated queries that can call one or more Web API functions. The returned results of these functions are then mapped into a knowledge graph defined with an ontology.

### 2.1 Federated data integration architecture implementation

To implement the federated data architecture solution at BASF, we relied on the SPARQL Micro-Service (SPARQL-MS) approach (Michel et al., 2019) to homogenise the syntax and access with the SPARQL query language. As a result, for each Web API function, we provided at least one SPARQL endpoint. The distinct SPARQL endpoints including the physical ones (over native RDF stores) are combined by writing and executing SPARQL 1.1 federated queries over the SPARQL federator endpoint as illustrated in Figure 1. In our prototype, the implementation of this federator was done by deploying a Virtuoso 7.2 server. This Virtuoso instance does not store any data from the original data sources such as OMA or STRING databases (see Fig. 1). We solely use the SPARQL 1.1 query engine, hence, in principle any other SPARQL 1.1. engine could be applied instead of Virtuoso.

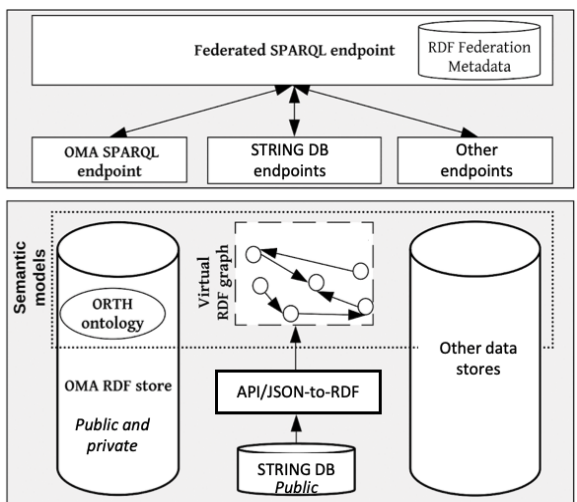

**Fig. 1.** A simplified view of the federation architecture in-use to interoperate Web APIs and SPARQL endpoints at BASF.

For simplicity, we call the graphs generated with SPARQL-MS the VKGs, nevertheless, they are quasi-virtual knowledge graphs (qVKGs) because the data are partially replicated during query time but it is not permanently stored. The data remain within the original data sources and accessed at query time. The partial replication occurs due to the fact of converting Web API call JSON responses into RDF triples, and temporarily storing the triples in an in-memory triple store, Corese-KGRAM[1]. However, the Web API calls return only a small portion of the whole dataset, consequently, the entire dataset is never fully converted into RDF triples. Solely a subset of the VKG is temporarily materialised, at query execution time, in order to answer the user query. This partial materialisation can potentially be used for caching. In this work, we applied the SPARQL-MS over the STRING database Web API functions[2]. Moreover, we have also applied SPARQL-MS to other BASF proprietary datasets, but due to confidentiality issues we will not describe them in this article. Further details of the SPARQL-MS deployment are available in Subsection 2.2. For the OMA database, it already provides a public SPARQL 1.1. endpoint over a native RDF store[3]. For implementing our prototype, we have chosen SPARQL-MS as an OBDA solution for Web APIs essentially because it is open-source and the ease of deployment.

### 2.2 Developed ontologies and mappings

As part of the solution, to build the VKGs, we reused existing public and internal BASF ontologies by also defining new terms when necessary. To address semantic and data schema heterogeneities, we sought to structure the data as homogeneously as possible by reusing the same graph patterns and ontologies already in use in private and public knowledge graph databases of our business interest such as the OMA database. For OMA, the core ontology is the Orthology (ORTH) ontology[4] version 2 (De Farias et al., 2017).

To deploy an OBDA approach, usually, we need three main elements: ontologies, mappings and data source access. **First**, we had to design an ontology to represent and describe the information retrieved from the STRING database WEB API functions. This ontology is composed of graph patterns and terms from several vocabularies such as ORTH v2 ontology, UniProt core ontology, and OBO foundry ontologies. The STRING DB ontology is fully described at http://purl.org/stringdb/ontology-doc. Although this ontology was specifically conceived for the STRING DB, it can be adapted and extended to represent similar protein-protein interaction data. **Second**, we defined the mappings to populate the designed ontology according to the Web API JSON responses. These mappings are defined as SPARQL construct queries for the SPARQL-MS approach. **Third**, we considered each Web API function call returning conceptually distinct results as a new data source, consequently, different data accesses were provided. By conceptually, we mean different JSON schemas used to structure the response, or semantically distinct results retrieved by modifying the function call parameters. For example, we use the same Web API function with different parameters to request distinct concepts such as functional and physical protein-protein interactions. In summary, all those three elements are available at https://purl.org/stringdb/sparql-git and used within SPARQL-MS. Similarly, we have defined these elements for deploying OBDA to the Web API functions over other BASF's proprietary datasets.

### 3 RESULTS : OMA AND STRING DATABASE USE CASE

We have evaluated our approach in terms of query execution time performance and retrieved results. In this evaluation, solely the public OMA and STRING databases were considered due to confidentiality issues. This evaluation was done with the end-to-end system: all queries were addressed to the federated SPARQL engine that dispatches parts of the query to their corresponding original data sources. The federated engine and SPARQL-MS services were deployed in the same machine: MacOS Big Sur, 3.3 GHz Dual-Core Intel Core i7 with 16 GB 2133MHz LPDDR3. Because of space constraints we summarise our evaluation in Table 1 and only describe in further details the query 8 (Q8). Each query was executed 10 times and caching results were disabled. More details about the other queries are provided at https://purl.org/stringdb/query-eval.

Q8 addresses the question: "What are the direct protein-protein functional interactions of a rice gene that is orthologous to the OMT2 wheat gene?". To answer this question, we need two databases: OMA, for getting the rice gene that is orthologous to the OMT2 wheat gene, and the STRING to get the protein-protein interactions of the protein encoded by this rice gene. All expected results were retrieved with a total mean time of around 1 second.

**Table 1.** Benchmark results of the federated data integration architecture implementation in terms of query execution time and retrieved results. The means and standard deviations related to the execution time of each query are defined in seconds.

| Query | Mean (s) | Std deviation | Retrieved results |
|---|---|---|---|
| Q1 | 0.95 | 0.51 | 10 |
| Q2 | 0.31 | 0.03 | 0 |
| Q3 | 0.74 | 0.12 | 26 |
| Q4 | 0.26 | 0.02 | 3 |
| Q5 | 0.32 | 0.01 | 1 |
| Q6 | 0.91 | 0.21 | 36 |
| Q7 | 0.37 | 0.07 | 0 |
| Q8 | 1.33 | 0.48 | 10 |

### 4 CONCLUSIONS

In this work, we demonstrated that current semantic web technologies allowed us to build a semantically enriched abstract layer with virtual knowledge graphs on top of the Web API functions that access the physical layer (i.e. original data sources). This abstract layer encapsulates and mitigates the complexity of the stored scientific data by accessing the needed information via the APIs. By doing so, we were also able to build a federated data integration prototype to interoperate Web APIs with native RDF stores that are public and private in an industry context. In this approach, the data integration is performed at query execution time mainly via conjunctive queries.

Finally, we highlight that an important area of future work will therefore be the comparison of SPARQL-MS with other commercial and noncommercial OBDA approaches over Web APIs such as Stardog[5]. Moreover, we can mention the design of federator tools enabling better usability, data source discovery, query plan definition, and indexing and caching capabilities. Given that SPARQL federators are not as mature as their SQL counterparts, major improvements and research are still required at the federation layer.

### ACKNOWLEDGEMENTS

The work was supported by a research agreement between BASF and SIB. We thank Stefanie De Bodt for contributing to make this collaboration possible. Furthermore, SIB is supported by the Swiss State Secretariat for Education, Research and Innovation (SERI).

---

[1] https://project.inria.fr/corese/corese-kgram/

[2] https://string-db.org/help/api/

[3] https://sparql.omabrowser.org

[4] https://qfo.github.io/OrthologyOntology/

[5] https://www.stardog.com/platform/connectors/rest/